\documentclass[10pt]{iopart}
\usepackage{graphicx}
\usepackage{amsfonts}
\expandafter\let\csname equation*\endcsname\relax
\expandafter\let\csname endequation*\endcsname\relax
\usepackage{amsmath}
\usepackage{amsbsy}
\usepackage{bbm}
\usepackage{dsfont}
\usepackage{color}
\usepackage{algorithm}
\usepackage{algpseudocode}
\usepackage{tikz}
\usetikzlibrary{positioning}
\usepackage{adjustbox}
\usepackage{bbm}
\eqnobysec 

\newcommand{\x}{\mathbf{x}}

\newcommand{\y}{\mathbf{y}}
\newcommand{\bt}{\mathbf{t}}
\newcommand{\ty}{\mathbf{t}^y}
\newcommand{\target}{\x}

\newcommand{\F}{\boldsymbol{F}}

\newcommand{\half}{\frac{1}{2}}

\newcommand{\DS}{\textcolor{blue}}
\newcommand{\cut}[1]{{}} 
\begin{document}
\title{Viral Load Inference in Non-Adaptive Pooled Testing}
\author{Mansoor Sheikh$^{\dag}$ and David Saad$^{\dag}$}
\address{$\dag$Non-linearity and Complexity Research Group, Aston University, Birmingham B4 7ET, United Kingdom}

\ead{d.saad@aston.ac.uk}
\date{\today}


\begin{abstract}
Medical diagnostic testing can be made significantly more efficient using pooled testing protocols. These typically require a sparse infection signal and use either binary or real-valued entries of $\mathcal{O}(1)$. However, existing methods do not allow for inferring viral loads which span many orders of magnitude. \cut{In this paper, we} We develop a message passing algorithm coupled with a PCR (Polymerase Chain Reaction) specific noise function to allow accurate inference of realistic viral load signals. This work is in the non-adaptive setting and could open the possibility of efficient screening where viral load determination is clinically important.


\end{abstract}


\section{Introduction}
Typically the infection status of a patient is determined by carrying out a single diagnostic test, which represents a poor use of resource in the low-prevalence case where most tests return negative. A well-studied improvement is the pooled testing concept~\cite{dorfman1943detection} which allows for the structured mixing of patient samples into groups or pools. By testing these mixtures (rather than the individual samples), the number of diagnostic tests required to determine each patient's infection status can be dramatically reduced. This can be beneficial where shortages of laboratory diagnostic equipment, raw testing materials and qualified staff occur. The two main approaches to pooled testing are the adaptive and non-adaptive protocols. In the former, tests are run sequentially, with information from the previous testing steps informing the next one~\cite{dorfman1943detection}\cite{mezard2011group}. However during the early spread of a virulent pathogen, laboratories are typically running at capacity and the logistics of adaptive pooled testing are not always feasible. In this paper, we focus on non-adaptive pooled testing which requires only one testing procedure to infer infection status. 

\begin{figure}
    \centering\includegraphics[trim={3cm 5cm 2cm 0cm} , clip , width=70mm,scale=0.5]{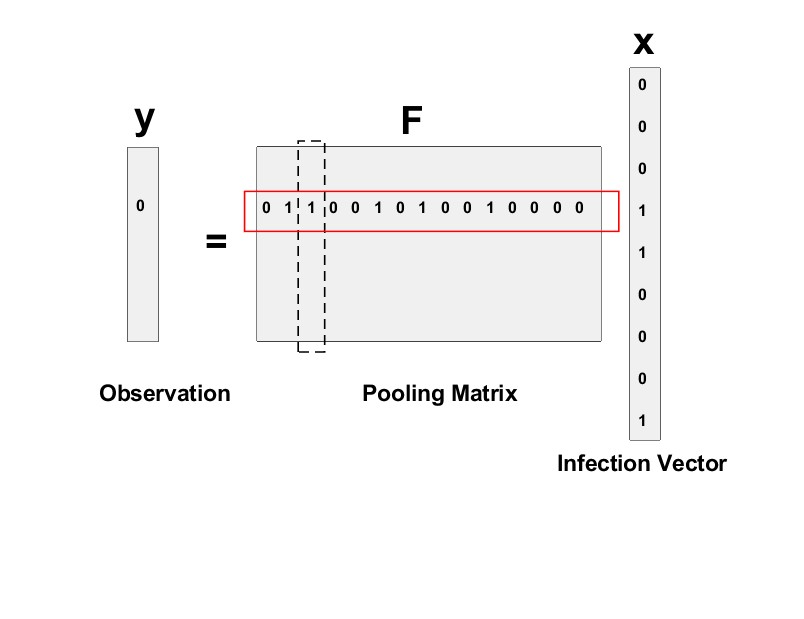}
    \caption{The inverse problem is to infer the $N \times 1$ infection vector, $\x$, given the $M \times 1$ observation vector, $\mathbf{y}$, and the $M \times N$ pooling matrix, $\F$.}
    \label{fig:matrix_mult}
\end{figure}

We consider the problem of recovering an unknown $N$-dimensional vector $\target$ representing the diagnostic status of $N$ patients where component values can be either binary or real.  We will introduce the concept using binary values. Combinations of samples are pooled according to the $M \times N$ pooling/measurement matrix, $\F$ (see Fig.~\ref{fig:matrix_mult}) where each row specifies which patient samples are included in each test. Matrix $\F$ can be constructed as either random or structured and is a known quantity in the inference problem. Each row of $\F$ can be thought of as probing/examining the unknown signal vector by taking a linear projection via a physical pooling of patient samples. The $M$ results are output as an $M$-dimensional vector, $\y$ and the aim is to solve the inverse problem of inferring $\target$ from known $\F$ and $\y$. A smaller measurement rate $\alpha = M/N$ corresponds to a more efficient pooled testing setup. 
The judicious design of an efficient pooling regime helps minimize $\alpha$. Typical approaches for real-valued variables $\target$ use pooling matrices with Gaussian random variable entries. Subsequent improvements include matrix designs from error-correcting codes~\cite{shental2020efficient} and from physics-inspired methods of crystal nucleation~\cite{zhang2013non}.

Individual infection status is determined from the pooled test measurements via a statistical or combinatorial inference procedure with the method used depending on the infection status represention e.g. binary status (typical in group testing) or real-valued viral loads (compressed sensing). These approaches to group testing have been investigated in the fields of computer science~\cite{ben2022recovery}, statistics~\cite{donoho2009message}\cite{ghosh2021compressed}, error-correcting codes~\cite{shental2020efficient} and statistical physics~\cite{zhang2013non}\cite{sakata2020bayesian}\cite{krzakala2012probabilistic}\cite{krzakala2012statistical}. Group testing typically assumes a sparse infection signal i.e. low disease prevalence. 

In some applications, the viral load can range from approximately $10^2 - 10^9$ copies per $mL$ e.g. in the case of SARS-CoV-2~\cite{pan2020viral}. This  makes inference problematic, especially for  inferring real-valued variables, and, to our knowledge, there is a lack of work dealing with this case. The problem stems from each pooled measurement containing a linear combination of zero, low, medium and high real values. Informally, the lower values will necessarily be "drowned out" by the high values in the pooling process. In this work, we do not assume that the signal entries are $\sim \mathcal{O}(1)$. First we ask the natural question of whether it is clinically relevant to ascertain the viral load of a sample rather than a simple absence or presence of the virus; the answer depends on the application:\\
\begin{itemize}
\item \emph{Viral load is clinically relevant but the efficiencies of pooled testing are not required.}
For chronic diseases such as HIV, a lower viral load means higher life expectancy \cite{kiplagat2022hiv}. Hence the success of treatment regimes is measured by viral load. For a previously infected patient, knowing the binary presence or absence does not matter because the virus will be there in some amount. In the case of HIV, the efficiencies of group testing are typically not required (unless it is done as community screening e.g.~\cite{krajden2014pooled}).
\item \emph{Widespread screening required but mostly the presence or absence of the disease is relevant}: 
In some cases knowing the viral load is less likely to affect the treatment or isolation regime (such as COVID or Hepatitis C).  Here low viral loads mean early infection stages but are less likely to affect the clinical decision. A possible \emph{caveat} to this is evidence that patients with high viral load are more efficient at spreading infection~\cite{beldomenico2020superspreaders}~\cite{liu2020viral}. Identifying and quarantining these individuals will have an outsized effect on viral spread. 
\item \emph{Both viral load and efficient screening are relevant}: Pooled testing can be used to estimate the amount of contaminants in food. Here it does matter how much salmonella there is in your chicken. There is also a clear commercial benefit for food companies to gain efficiencies from pooled testing.
\item \emph{Inferring the stage of outbreak}: The distribution of $C_t$ values in a population is related to the stage of the viral outbreak \cite{hay2021estimating}\cite{singanayagam2022community}. By inferring $C_t$ values, the method described in the current paper provides a resource-efficient method of estimating the stage of infection and potentially the viral reproduction number, $R_0$.
\end{itemize}
A naive approach would be to run a compressed sensing algorithm to recover the viral load signal which has a high dynamic range. Compressed sensing is a theoretically and empirically accurate signal recovery scheme \cite{1228344}\cite{donoho2009message}\cite{krzakala2012statistical} where the typical error is orders of magnitude smaller than the signal magnitude. However\DS{,} this results in small signal components becoming indistinguishable from the noise. Synthetic studies conventionally sample the non-zero viral loads from a uniform distribution e.g. $x \sim \mathcal{U}(2^0, 2^{15})$ in \cite{ghosh2021compressed}. Here approximately $90\%$ of the samples lie in the range $(2^{12}, 2^{15})$, which is above the typical noise level in compressing sensing algorithms, leading to flattering accuracy. The claimed high dynamic range is \emph{probably} not that high. This argument is applicable to a lesser extent in \cite{cohen2021multi} where viral loads are sampled \emph{uniformly} from $[0,1000]$ but discretization results in $70\%$ of infected samples being in the mid $[300,700)$ and high $[700,1000]$ categories. In the current paper, we sample uniformly from a logarithmic range i.e. equal number of samples from each of $2^0, 2^1, 2^2, \ldots $ which is more realistic and relevant. This ensures there is truly a high dynamic signal range but results in a materially harder problem.

\section{Model}
\subsection{Standard combination protocols}
\label{sec:standard_setup}

There are two typical combination protocols associated with the inverse problem of Fig.~\ref{fig:matrix_mult}:
\begin{enumerate}
    \item Standard matrix multiplication of $\F$ and $\target$ e.g.~\cite{zhang2013non}
    \begin{equation}
    \label{eq:standard_mm}
    \tilde{y}_\mu= \sum_{i=1}^N F_{\mu i} x_i + \xi_\mu
    \end{equation}
    where $\xi_\mu$ represents Gaussian noise associated with test $\mu$, $x_i$ the individual loads, $F_{\mu i}$ the mixing matrix and $\tilde{y}_\mu$ the noisy test result (a corrupted version of  $\mathbf{y}$). This is termed the \emph{linear estimation problem}~\cite{rangan2011generalized}
    \item The binary testing regime uses a logical sum e.g.\cite{sakata2020bayesian}. If one patient is infected, the test output is infected.
    \begin{equation}
    	       \tilde{y}_\mu= C \left(
    	\bigvee_{i=1}^N
    	F_{\mu i} ~x_i 
    	\right)
       \cut{ t_\mu^y = C \left(
        \bigvee_{i=1}^N
        F_{\mu i} ~t_i 
        \right)}
    \end{equation}

where $x_i$ the individual presence/absence of the disease,  $\tilde{y}_\mu$ the noisy test result and $C(\ldots)$ is a probabilistic function incorporating measurement noise such as false positive and false negative rates. Since each patient participates in multiple tests (so-called overlapping tests), the accuracy achieved can improve upon the device error settings~\cite{sakata2020bayesian}.
\end{enumerate}

\subsection{PCR specific notation}
\label{sec:setup}
\begin{figure}[h]
\centering\includegraphics[trim={1cm 0.7cm 0cm 0cm} , clip , width=80mm,scale=0.5]{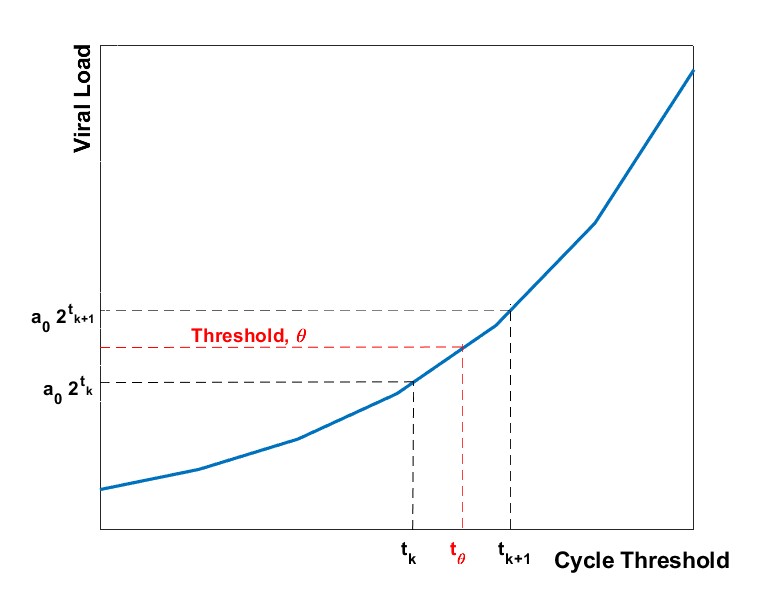}
    \caption{This schematic plot shows the doubling process central to PCR devices. Since detection is at integer values, the observed $C_t$ reading is "rounded-up" from the exact solution to $a_0 2^{t^{\theta}} = \theta$ where $a_0$ is a random variable which represents the initial viral load of the sample (notice that $t^{\theta}$ is not an integer). We do not observe $a_0$ directly but only via an integer valued $\lceil t^{\theta} \rceil$. Reading off from curve $a_0 2^{t_k} < \theta \leq a_0 2^{t_{k+1}}$. Re-arranging $\theta 2^{-t_{k+1}} \leq a_0 < \theta 2^{-t_k}$. }
    \label{fig:doubling}
\end{figure}

In this paper, we focus on amplification methods such those used in the Polymerase Chain Reaction (PCR) device.  The initial patient sample is repeatedly heated and cooled to encourage a doubling of the viral RNA (see Fig.~\ref{fig:doubling} for a schematic representation). Hence the initial viral load is amplified. A marker is added which fluoresces when attached to a specific section of the virus. Once sufficient fluorescence is detected with an optical device, the cycling is stopped and the resulting cycle number recorded. If no fluorescence is detected after an upper limit of cycles, typically 40, it is determined that no virus was originally present. The upper cycle threshold ($C_t$) limit is determined by the \emph{limit of detection} parameter of the PCR device. A $C_t$ value of 20 typically signifies a high viral load. We now define the notation related to the PCR protocol. 
\begin{itemize}
    \item $\theta=$ threshold for detection of fluorescence. This is typically measured in number of viral copies per mL of transport media and is assumed constant for a given PCR device.
    \item $a_i=$ initial viral load  (number of viral copies per mL). The higher the initial viral load $a_i$, the lower the number of amplification cycles needed to detect the virus.
    \item $t_i=$ number of amplification cycles required to detect fluorescence if sample $i$ is tested individually. Values of $t_i$ are read off as integers rather than the "exact" real valued solution to $\theta = a_i 2^{t_i}$.
    \item $t_{\mu}^y=$ number of amplification cycles required to detect fluorescence for the pooled test $\mu$. 
\end{itemize}

Given the nature of the PCR doubling cycle, the linear projections described in (\ref{sec:standard_setup}) no longer hold. This is because the viral loads are averaged in the pooling process but these viral loads are only observed via the integer cycle number. A new combination protocol is formulated in the next section.
\subsection{Simple mixing example}
To gain intuition for the idiosyncrasy of PCR mixing, consider two patients $i$ and $j$ where $\theta = a_i 2^{t^{\theta}_i}$ and $\theta = a_j 2^{t^{\theta}_j}$ represent each patient being tested individually (note that $t^{\theta}$ are not integers). The aim is to recover $t_i$ and $t_j$ from the possibly noisy measurement vector $\tilde{\y}$. If these samples are combined in an equal ratio in test $\mu$, the resulting viral load, which is never directly observed, is $\half (a_i+a_j) = \frac{\theta}{2} (2^{-t^{\theta}_i}+2^{-t^{\theta}_j})$. The corresponding measurement cycle number, which we label $t_{\mu}^y$ is given by the relationship $2^{-t_{\mu}^{\theta}} = \half (2^{-t^{\theta}_i}+2^{-t^{\theta}_j})$ where the $\theta$ parameters cancel. To illustrate numerically, suppose the initial viral loads are $a_i=2^5=32$ and $a_j=2^6=64$ and the detection threshold $\theta=2^9=512$. We find $t^{\theta}_i=t_i=4$ and $t^{\theta}_j=t_j=3$ noting the inverse relationship between $t^{\theta}_i$ and $a_i$. For samples pooled in an equal ratio, the viral load per unit volume is $\half (32+64) = 48$ leading to $t_{\mu}^{\theta} \approx 3.42$, a non-integer value between $t^{\theta}_i$ and $t^{\theta}_j$. The PCR device will return a measurement $C_t$ value of $\lceil t_{\mu}^{\theta} \rceil = 4$. This measurement discretization makes the inference problem more difficult and less accurate.

\subsection{General mixing example}
\label{sec:mixing_example}
If equal quantities of $K$ samples are pooled (where $K < N$), the corresponding mixing relationship is 
	\begin{equation}
\label{eq:K_samples}
  2^{-t_{\mu}^{\theta}} = \frac{1}{K} \sum\limits_{k=1}^K 2^{-t^{\theta}_{\mu k}}
\end{equation}
where $t^{\theta}_{\mu k}$ represents the real-valued threshold of element $k$ in the mix and $t_{\mu}^{\theta}$ the  real-valued threshold for test $\mu$.

\subsection{Signal sparsity assumption}
In traditional testing, we perform $N$ diagnostic tests for $N$ patients. Attempts to reduce the number of tests i.e. $M < N$ will lead to an ill-defined system of equations in Fig.~\ref{fig:matrix_mult} with an infinite number of solutions. The constraint required to solve the problem typically relies on a sparsity assumption (in some suitable basis) i.e. a majority of zero entries corresponding to non-infected patients. In the present case, uninfected patients correspond to a cycle number equal to the upper limit. In typical compressed sensing scenarios, a sparse prior is defined via the signal density $\rho$ using a Bernoulli-Gauss distribution~\cite{krzakala2012probabilistic} such as    $p(\bt) = \prod\limits_{i=1}^N \left[
(1-\rho) ~\delta(t_i) + \rho ~ \mathcal{N}(\mu, \sigma^2)
\right]$ and i.i.d. signal components. In this paper, the variables used are all integers, corresponding to the PCR cycles where infection can be first detected. They range from the lowest number of cycles $L$ to the highest $U$. In the absence of additional  information we assume the non-zero signal entries take uniformly distributed values in the state space $\mathcal{S} \in \{L, L+1, \ldots, U\}$, according to 
\begin{equation}
\label{eq:sparse}
    p(\bt) = \prod\limits_{i=1}^N \left\{ (1-\rho) \delta_{t_i,U} + \rho ~ \Theta[t_i- L]~\Theta[(U-1)-t_i]
    \right\}    
\end{equation}
where $\Theta(x)$ is the Heaviside step function and $U$ represents no viral load/not infected.

An alternative approach, used in~\cite{ben2022recovery}, is to discretize the entire signal range $\mathcal{S} \in \{L, \ldots, U\}$ into low, medium, high and non-infected ranges corresponding to clinically relevant ranges of low, mild and high infection status. This simplification simplifies the inference problem at the expense of accuracy.
%
\section{Message passing}
We aim to recover the  $N$-dimensional vector $\bt$ where each component $t_i \in \mathcal{S}$. The viral load variables only participate through the mixing relationship described in Sec.~\ref{sec:mixing_example}. The inference problem can be represented by a bipartite factor graph (see Fig.~\ref{fig:bipartite_graph}) and since the graph is sparsely connected, inference can be efficiently achieved using a message passing algorithm, whereby conditional probabilities are iteratively exchanged between factors and variables until they converge to provide pseudo marginal posterior probabilities for the individual variables. Message passing methods are exact on trees but offer approximate solutions on loopy graphs~\cite{MFbook2001}. Their use in the context of group testing has been studied previously but mostly in the case of binary infection status to our knowledge~\cite{sakata2020bayesian,mezard2008group}. While preparing the manuscript, we came across the paper~\cite{ben2022recovery} which utilizes a message passing algorithm for inference of a real-valued signal. However, there is a distinct difference between our approach and the one of ~\cite{ben2022recovery}, that is based on the iterative removal of values found and uses a very small number of categories, arguably making the problem easier.

Messages exchanged between factors and variables represent a closed set of equations of messages from factors to variables  $p(f_{\mu}|t_i) \equiv m_{\mu \to i}(t_i) $ and from variables to factors $p(t_i |\{\mathbf{f}\}_{\backslash \mu}) \equiv m_{i \to \mu}(t_i)$, detailed below.


\cut{\begin{figure}
\label{fig:bipartite_graph}
\hspace{3cm} 
\begin{tikzpicture}[node distance=1.00cm and 0.5cm,>=latex, rotate=90]
    \node[draw, circle] (x1) {$x_1$};
    \node[draw, circle, right=of x1] (x2) {$x_2$};
    \node[draw, circle, right=of x2] (x3) {$x_3$};
    \node[right=of x3] (xdots1) {$\dots$};
    \node[draw, circle, right=of xdots1] (xi) {$x_i$};
    \node[right=of xi] (xdots) {$\dots$};
    \node[draw, circle, right=of xdots] (xn) {$x_j$};
    \node[draw, circle, right=of xn] (xnn) {$x_n$};
    
    \node[draw, rectangle, above=of x2] (f1) {$f_1$};
    \node[draw, rectangle, above=of x3] (f2) {$f_2$};
    \node[right=of f2] (fdots1) {$\dots$};
    \node[draw, rectangle, above=of xi] (f3) {$f_{\mu}$};
    \node[right=of f3] (fdots2) {$\dots$};
    \node[draw, rectangle, above=of xn] (fm) {$f_m$};

    \draw[->] (f1) -- (x1) node[midway, above] {$\mathbf{m_{f_1 \to x_1}}$};
    \draw[->] (xnn) -- (fm) node[midway, above] {$\mathbf{m_{x_n \to f_m}}$};

    \draw[-] (f1) -- (x2);
    \draw[-] (f1) -- (xi);
    \draw[-] (f2) -- (x1);
    \draw[-] (f2) -- (x3);
    \draw[-] (f2) -- (xn);
    \draw[-] (f3) -- (x2);
    \draw[-] (f3) -- (xi);
    \draw[-] (f3) -- (xnn);
    \draw[-] (fm) -- (x3);
    \draw[-] (fm) -- (xn);
    \draw[-] (fm) -- (xi);
\end{tikzpicture}
\caption{Random variables (circles) represent each of the $N$ patients tested. Each factor node (squares) represents a test/measurement $\mu$. The degree of the factors nodes is $K$ and of the variable nodes is $L$}
\end{figure}
}
\begin{center}
\begin{figure}
	\label{fig:bipartite_graph}
	\begin{adjustbox}{center}
	\hspace{3cm} 
	\begin{tikzpicture}[node distance=1.00cm and 0.5cm,>=latex, rotate=90]
		\node[draw, circle] (x1) {$t_1$};
		\node[draw, circle, right=of x1] (x2) {$t_2$};
		\node[draw, circle, right=of x2] (x3) {$t_3$};
		\node[right=of x3] (xdots1) {$\dots$};
		\node[draw, circle, right=of xdots1] (xi) {$t_i$};
		\node[right=of xi] (xdots) {$\dots$};
		\node[draw, circle, right=of xdots] (xn) {$t_j$};
		\node[draw, circle, right=of xn] (xnn) {$t_n$};
		
		\node[draw, rectangle, above=of x2] (f1) {$f_1$};
		\node[draw, rectangle, above=of x3] (f2) {$f_2$};
		\node[right=of f2] (fdots1) {$\dots$};
		\node[draw, rectangle, above=of xi] (f3) {$f_{\mu}$};
		\node[right=of f3] (fdots2) {$\dots$};
		\node[draw, rectangle, above=of xn] (fm) {$f_m$};
		
		\draw[->] (f1) -- (x1) node[midway, above] {$\mathbf{m_{f_1 \to t_1}}$};
		\draw[->] (xnn) -- (fm) node[midway, above] {$\mathbf{m_{t_n \to f_m}}$};
		
		\draw[-] (f1) -- (x2);
		\draw[-] (f1) -- (xi);
		\draw[-] (f2) -- (x1);
		\draw[-] (f2) -- (x3);
		\draw[-] (f2) -- (xn);
		\draw[-] (f3) -- (x2);
		\draw[-] (f3) -- (xi);
		\draw[-] (f3) -- (xnn);
		\draw[-] (fm) -- (x3);
		\draw[-] (fm) -- (xn);
		\draw[-] (fm) -- (xi);
	\end{tikzpicture}
	\end{adjustbox}
	\caption{Random variables (circles) represent each of the $N$ patients tested. Each factor node (squares) represents the compatibility of the test/measurement $\mu$ with the relevant patient variable values. The degree of the factors nodes is $K$ and of the variable nodes is $L$}
\end{figure}
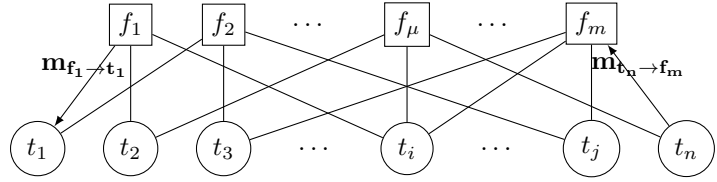
\end{center}

\subsection{Factor-to-variable messages (PCR noise)}
Here we denote the noise model as $\phi(\bt, \ty)$ where $\bt \in \{t_1, t_2, \ldots, t_K\}$. The mechanism needed for the discretization inherent in the PCR measurement process is developed in Sec.~\ref{sec:develop_noise_model}.
\begin{equation}
\begin{split}
\label{eq:factor_to_variable_PCR}
    &m_{\mu \to i}^{(\tau+1)}(t_i) \propto \sum\limits_{\bt \backslash t_i} \phi(\bt, \ty)
    \prod\limits_{j \in \partial \mu \backslash i}^N m_{j \to \mu}^{(\tau)}(t_j)
\end{split}
\end{equation}
The superscript $\tau$ denotes the iteration step and the notation $\partial \mu \backslash i$ refers to all variables connected to factor $\mu$ except $i$. Each message is an $|\mathcal{S}|$-dimensional column vector with entries being real numbers between 0 and 1 representing the state probability. The summation is over all possible states of the vector $\bt$ excluding $t_i$. This is the main computational bottleneck and does not scale well with $|\mathcal{S}|$ (state space size) or $K$ (factor node degree). This difficulty is reduced in~\cite{ben2022recovery} by assuming a state space of none, low, medium and high viral loads. For real-valued signals, the messages can be approximated by their means and variances leading to approximate message passing protocols~\cite{krzakala2012statistical,1228344}.

\subsection{Variable-to-factor messages}
\begin{equation}
\label{eq:variable_to_factor1}
    m_{i \to \mu}^{(\tau+1)}(t_i) \propto \Big\{ (1-\rho) \delta_{t_i,U} + \rho ~ \Theta[t_i- L]~\Theta[(U-1)-t_i]
    \Big\}
    \prod\limits_{\gamma \in \partial i \backslash \mu} m_{\gamma \to i}^{(\tau)}(t_i)
\end{equation}
where $\delta_{ij}$ is the Kronecker delta function and the notation $\partial i\backslash \mu$ refers to all  factors connected to variable $i$ except $\mu$. For both factor-to-variable and variable-to-factor messages, the constants of proportionality can be calculated by normalisation e.g. $\sum\limits_{t_i \in \mathcal{S}} m_{i \to \mu}(t_i) = 1$. 

\subsection{Marginal Probabilities} Once the messages~\eqref{eq:factor_to_variable_PCR}\eqref{eq:variable_to_factor1} have been iterated to convergence, the marginal probabilities for each variable/patient can be calculated: 
\begin{equation}
\label{eq:calc_marginals}
    p(t_i) \propto 
    \Big\{ (1-\rho) \delta_{t_i,U} + \rho ~ \Theta[t_i- L]~\Theta[(U-1)-t_i]
    \Big\}
    \prod\limits_{\mu \in \partial i }  m_{\mu \to i}(t_i) \\
\end{equation}
The inferred $C_t$ value corresponds to the message component with highest probability. It is an approximation in our case since the bipartite graph contains loops.

\section{Developing the noise model}
\label{sec:develop_noise_model}
The two types of noise present in the PCR mechanism are discretization noise and physical measurement noise. We will address the former in this section and the latter in Sec.~\ref{sec:accuracy_noise}. Discretization noise is present since $C_t$ values are recorded as integers rather than decimals and requires careful treatment in the mathematical modelling setup. 

\subsection{No measurement noise}
In the case where the real cycle values are applied (denoted by the $\theta$ superscript) Eq.~(\ref{eq:K_samples}) holds, and consequently
	\begin{equation}
		\label{eq:no_noise_case}
		\phi(\bt, \ty) = \prod\limits_{\mu=1}^M~ \delta\left(
		\frac{1}{K} \sum\limits_{k=1}^K 2^{t^{\theta}_{\mu}-t^{\theta}_{\mu k}} - 1
		\right)
	\end{equation}
However, discretization to the next higher integer cycle value creates a difference between the test cycle integer value and those of the individual samples. The modelling of this difference is a challenge and an appropriate noise model has to be used.
\cut{We start with the noiseless case:
\begin{equation}
\label{eq:no_noise_case}
    \phi(\bt, \ty) = \prod\limits_{\mu=1}^M~ \delta\left(
    \frac{1}{K} \sum\limits_{k=1}^K 2^{t_{\mu}^y-t_{\mu k}} - 1
    \right)
\end{equation}
where $t_{\mu}^y$ represents the $C_t$ value for the $\mu$-th pooled test and $t_{\mu k}$ is the $C_t$ value for the $k$-th non-zero entry of the $\mu$-th row of $\F$. This delta function approach does not work since PCR outputs are integer $C_t$ values. 
}

\subsection{Gaussian measurement noise}
One of the simplest models to accommodate these differences is to assume measurement errors follow a Gaussian distribution incorporating~\eqref{eq:K_samples} and the discretization (see~\cite{ghosh2021compressed}). We will replace the ``true'' but unknown real cycle values $t^{\theta}_{\mu}$ and $t^{\theta}_{\mu k}$ by the integer variables $t_{\mu}^y$ and $t_{\mu k}$, respectively. We expect that $t_{\mu}^y=\lceil t^{\theta}_{\mu}\rceil$ and $t_{\mu k}=\lceil t^{\theta}_{\mu k}\rceil$.
\begin{equation}
\label{eq:noisy_case}
\phi(\bt, \ty) = \prod\limits_{\mu=1}^M \frac{1}{\sqrt{2\pi \Delta_\mu}} \exp\left[-\frac{1}{2\Delta_\mu}
\left(
\frac{1}{K} \sum\limits_{k=1}^K 2^{t_{\mu}^y-t_{\mu k} } - 1
\right)^2\right]
\end{equation}
where $\Delta_\mu$ is the variance in the noise measurement.

We will consider two different noise distributions $\phi(\bt, \ty)$ to "filter out" unreasonable combinations of $\bt$ and $\ty$ from the factor-to-variable messages in~\eqref{eq:factor_to_variable_PCR}. The first is a simple binary function in Sec.~\ref{sec:step_function} and the second uses the overlap between distributions to weight message products in Sec.~\ref{sec:distribution_overlap}.

\subsection{Step-function distribution}
\label{sec:step_function}
First consider the measure $d$ inspired by \eqref{eq:K_samples} but using integer rather than real values.
\begin{equation}
    d = \left(\frac{1}{K} \sum\limits_{k=1}^K 2^{-t_k}\right) - 2^{-t^y} = X - Y
\end{equation}
where we will use short-hands $X \equiv \frac{1}{K} \sum\limits_{k=1}^K 2^{-t_k}$ and $Y \equiv 2^{-t^y}$ which are fixed for a given combination. The uncertainty in both random variables $X$ and $Y$ can be represented by uniform distributions $\mathcal{U}(X,2X)$ and $\mathcal{U}(Y,2Y)$ leading to an inequality:
\begin{equation}
\label{eq:step_inequality}
    \frac{1}{2}2^{-t^y} \leq \frac{1}{K} \sum\limits_{k=1}^K 2^{-t_k} \leq 2 \cdot 2^{-t^y}
\end{equation}
Hence our first approximation for $\phi(\bt, \ty)$ can be written: 
\begin{equation}
   \phi(\bt, \ty) = 
   \begin{cases} 
      1 & \text{if } -\frac{1}{2}2^{-t^y}\le d \le 2^{-t^y} \\
      0 & \text{otherwise.}
   \end{cases}
\end{equation}
Notice this function does not need to be normalized since the messages are normalized in the message passing iterations. Clearly $d=0$ is in the interval as expected. The message passing summation \eqref{eq:factor_to_variable_PCR} includes all combinations of $\{t_k\}$ for a given $t_y$ and the role of $\phi(\bt, \ty)$ is to exclude implausible combinations. This simple binary approach has the benefit of materially reducing the number of summands in each message passing iteration speeding up implementation.

\subsection{Distribution Overlap}
\label{sec:distribution_overlap}
Rather than a binary function, consider a function $\phi(\bt, \ty)$ which weights different combinations of $\{t_k\}$ and $t_y$ depending on the overlap of the distributions $\mathcal{U}(X,2X)$ and $\mathcal{U}(Y,2Y)$. For the case $X<Y$:
\begin{equation}
\phi(\bt, \ty) = 
\begin{cases} 
0 & \text{if } X \leq \frac{1}{2}Y \\ \\
\frac{2X-Y}{Y} & \text{if } \frac{1}{2}Y \leq X \leq Y \\ \\
\frac{2Y-X}{Y} & \text{if } Y \leq X \leq 2Y \\ \\
0 & \text{if } X > 2Y
   \end{cases}
\end{equation}
This noise distribution, which has the same support as \eqref{eq:step_inequality}, is used in the numerical experiments of Sec.~\ref{sec:numerical_experiments}. 

\subsection{Further refinement}
Using a uniform distribution for $ 2^{-t^y}$ is plausible but the distribution for $X=\frac{1}{K}\sum_k 2^{-t_k}$ should properly account for each $t_k$ chosen at random from $\{L, \ldots, U\}$. The probability density function has been derived in ~\cite{bradley2002distribution}\cite{sadooghi2009distribution}. The cumulative distribution function for $X$, derived in~\cite{buonocore2009note}, can be used to calculate the overlap with $\mathcal{U}( 2^{-t^y} , 2 \cdot  2^{-t^y})$. This calculation is not implemented due to the increased computational cost required due to the nested summations. 


\subsection{Alternative treatments}
Previous research has considered the PCR noise function but primarily in the case of binary variables (infected/non-infected). The traditional noise measures of false positive and false negative rates are comprehensively treated in~\cite{sakata2020bayesian} where it is shown that pooled testing can overcome technical device limitations since each patient participates in multiple tests. Noise in $C_t$ values and fluoresence thresholds are dealt with in~\cite{ghosh2021compressed} but with a different signal reconstruction algorithm and the discretization effect is modelled as Gaussian rather than uniform. We have avoided the need for calibration of the threshold $\theta$ and mapping between amplified viral load and fluoresence in~\cite{ghosh2021compressed} with our functional form of $\phi(\bt, \ty)$. An alternative approach is to learn the translation function between viral loads and discrete measurement classes using supervised learning methods such as a neural network~\cite{ben2022recovery}. While our approach splits this up into mixing~\eqref{eq:K_samples} and noise probability functions,~\cite{ben2022recovery} may be able to learn from previous PCR results \emph{if the ground truth properties are  known}.


\section{Numerical Experiments}
\label{sec:numerical_experiments}
\subsection{Implementation}
The problem is defined by initializing parameters $\rho, M, N, K$ and state space $\mathcal{S}$. Of course, the signal sparsity $\rho$ is not \emph{a priori} known for a given batch of samples but could be estimated by either the Expectation-Maximization~\cite{sakata2020bayesian}\cite{krzakala2012probabilistic} or the Expectation Propagation~\cite{braunstein2020compressed} methods. An approximate but practical approach is to take a recent average disease prevalence value as an estimate of $\rho$. This parameter is used to set the number of observations $M$ in the pooling matrix. 

We now carefully describe the protocol for generating the ground truth signal $\bt$ in our numerical experiments. This is important as it influences our subsequent choice of noise function boundary. \emph{Decimal} values are sampled for infected and non-infected
patients from uniform distributions $\mathcal{U}(L-1,U-1)$ and $\mathcal{U}(U-1,U)$ respectively and combined using~\eqref{eq:sparse}. The \emph{decimal} measurement vector $t_y^{\theta}$ is calculated \emph{in silico} using the PCR mixing protocol~\eqref{eq:K_samples}. Finally an \emph{integer} measurement value is obtained by rounding up $t_y^{\theta}$ to the nearest integer. Additional noise is added (see Sec.~\ref{sec:accuracy_noise}) in some scenarios as described below. The discretized version of $\bt$ is stored for calibration purposes once the inference is complete. A different (unphysical) protocol could be considered which starts from \emph{integer} values of infected patients. This would reduce uncertainty in the noise function thus providing more flattering accuracy statistics but is not used in our experiments since it unrealistic.

The random binary measurement matrix $\F$ defines which samples are included in each test and is typically defined by the degree distribution of the variable and factor nodes of the bipartite graph (Fig.~\ref{fig:bipartite_graph}). Choices include random-random or random-Poisson~\cite{mezard2008group}. In this paper, we follow~\cite{sakata2020bayesian} in using random-random matrices such that each row has $K$ non-zero entries and each column has $L=KN/M$ non-zero entries. In other words, the factor nodes have a constant degree $K$ and the variable nodes a constant degree $L$. The values used in our setup result in a sparse $\F$ matrix. Our message passing algorithm could also be combined with the structured matrix approach~\cite{zhang2013non,krzakala2012probabilistic,angelini2012compressed}. 

Given knowledge of $\F$ and $\y$, the message passing equations~\eqref{eq:factor_to_variable_PCR},~\eqref{eq:variable_to_factor1} are iterated until convergence. The message passing equations search for a fixed point to the dynamical system of messages, from which posterior variable values can be inferred using~\eqref{eq:calc_marginals}. Typical of iterative problems, our implementation utilises a damping factor (in our case, set to $0.01$) to help convergence. In addition, the sum of squared differences between all messages is chosen as a metric to determine convergence. The threshold was chosen empirically as $10^{-5}$ for the noiseless cases. 
\cut{The required marginal posterior values are calculated from the converged messages using~\eqref{eq:calc_marginals}.} This synthetic setup can therefore be  used to test the accuracy of signal recovery. 

\subsection{Easy Algorithm}
In problems where only binary infection status is considered,  \emph{sure variables} of certain states can be determined using simple logical arguments without resorting to more complex inference methods. These are termed \emph{Combinatorial Basis Pursuit (CBP)} or \emph{Combinatorial Orthogonal Matching Pursuit (COMP)} in computer science implementations of group testing~\cite{chan2011non} and are described in an \emph{Easy Algorithm}~\cite{mezard2008group}. Briefly, they identify certain (sure) negative patients since they participate in a negative test result. Further, sure positive samples can be found as those who participate in positive tests alongside sure negative samples. These variables are then removed reducing the problem size. One drawback of this approach is the inability to infer \emph{undetermined variables}~\cite{mezard2008group}, the inability to use prior knowledge and to exploit the power of probabilistic inference in the inevitable presence of measurement noise.

In the present case, where the state space $\mathcal{S}$ represents $C_t$ values, a non-infected test measurement result, $t_{\mu}^y = U$, does not guarantee all samples are non-infected (even in the noiseless case) e.g. true $C_t$ values of $\{U-1, U, U, U, U\}$ will result in $t_{\mu}^y = U$. This is due to the peculiarity of the PCR mixing protocol~\eqref{eq:K_samples} and the discretization of $C_t$ values. The issue can be seen clearly for the low viral load values (equivalently high $C_t$ values) in Fig.~\ref{fig:confusion_matrix}. Hence the \emph{Easy Algorithm} cannot be used in our work, either as a pre-processing step or as a comparison baseline for our message passing results. In fact it is arguably impractical to any scenario that includes a measurement noise.

\subsection{Uncertain Inference}
\label{sec:uncertain}
Recall that each message consists of $|\mathcal{S}|$ components representing the probability of each state from $L$ to $U$. To be clear, each state of $\mathcal{S}$ represents a doubling of the viral load from the preceeding state on a logarithmic scale (base 2). Our Maximum A Posteriori \emph{(MAP)} inference protocol uses the highest probability state as our diagnosis. However, for some samples, two states could have similar high probabilities implying uncertain inference. A \emph{computational} improvement could be envisaged, using decimation, where, upon convergence, the less ambiguous results are fixed to a standard basis vector e.g. $(0,1,0,\ldots,0)^T$ and the message passing algorithm continues with a smaller convergence threshold. Note, the "certain" samples are simply the complement of the uncertain samples.  This intervention in the algorithm dynamics did not meaningfully change the results in our test implementations. A possible \emph{clinical} protocol would be to physically re-test this small number of individuals with ambiguous results, but this has its own operational and medical implications.

\subsection{Motivating Example}
\label{sec:motivate}
\begin{figure}
    \centering    \includegraphics[trim={0cm 0cm 0cm 0cm} , clip , width=80mm,scale=0.5]{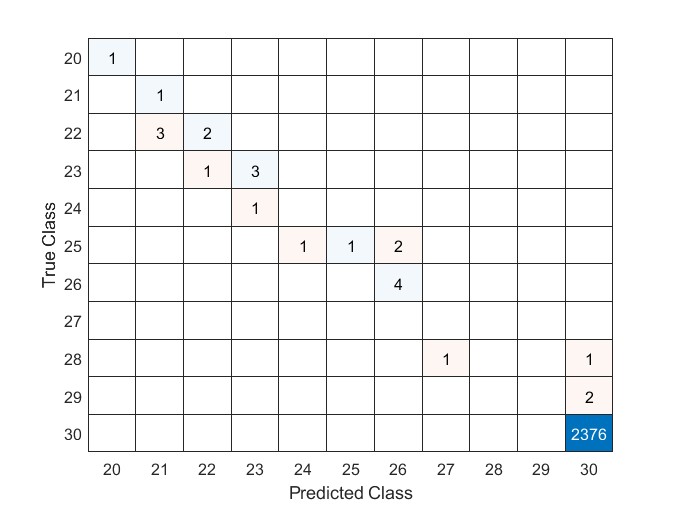}
    \caption{Confusion matrix representing a single problem instance, described in Sec.~\ref{sec:motivate}, using synthetic data: $M=800,~N=2400,~K=6,~\rho=0.01, ~\eta=0$ and $\mathcal{S} \in \{20,\ldots,30\}$.  The noise function $\phi(\bt, t_y)$ is taken as described in Sec.~\ref{sec:distribution_overlap}, producing an accuracy of 2388/2400 or 99.5\% (compared to 99.0\% for a always-negative classifier). The false negative rate of $3/24=12.5\%$ reduces as the state space $\mathcal{S}$ increases (see Fig.~\ref{fig:sens_vs_S}).} 
    \label{fig:confusion_matrix}
\end{figure}

To illustrate the discretization problem central to PCR testing, Fig.~\ref{fig:confusion_matrix} shows the confusion matrix for the noiseless inference problem of $N=2400$ individual samples, $M=800$ observations, $K=6$ patients in each group, $\rho=0.01$ prevalence and $\mathcal{S} \in \{20, \ldots, 30\}$. As a guideline, the expected number of tests required in adaptive {\em{binary (yes/no only)}} Dorfmann tests is approximately $2\sqrt{\rho}N = 480$ (see~\cite{mezard2008group}).
Predicting a $C_t$ value one away from its true value is clinically acceptable leading to the formulation of two accuracy measures $C_0$ and $C_{\pm 1}$ corresponding to a tolerance of zero and $\pm 1$ $C_t$ value, respectively. However, given the imbalanced classes in the signal, this metric is found not to be suitable to our task since simply predicting no infection results in high accuracy. 

\begin{figure}
    \centering    \includegraphics[ width=80mm,scale=0.5]{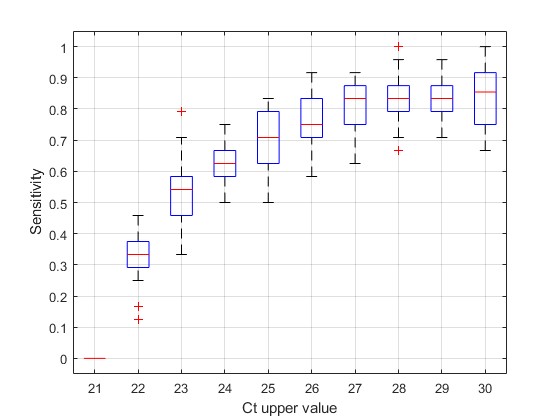}
    \caption{Plot of Sensitivity (True Positive Rate = TP/P) against the largest $C_t$ value. The lower $C_t$ value is fixed at 20. The intuition from our motivating example in Fig.~\ref{fig:confusion_matrix} is false negatives occur for $C_t$ values close to the maximum. This plot confirms false negatives decrease as the range size become larger the number of values per variable is higher. The zero sensitivity value at $C_t=21$ corresponds to all $24$ positive samples being falsely classified as negative. Parameters $\rho=0.01, ~\alpha = M/N = 800/2400$ are fixed. Changing a true to a false positive results in a sensitivity change of $1/24 \approx 4\%$. Simulations are run 30 times. Red lines display the median value, the boxes cover the interquartile range and the whiskers show the extreme values. }
\label{fig:sens_vs_S}
\end{figure}

For the remainder of this paper, we  will focus on the clinically relevant metric of sensitivity (the ratio of true positives to positive samples) and use a state space comprising $\mathcal{S} \in \{20,\ldots,30\}$. The relationship between sensitivity and $|\mathcal{S}|$ is investigated in Fig.~\ref{fig:sens_vs_S}.


\subsection{Accuracy of inference procedure}
\label{sec:accuracy}
\begin{figure}
    \centering    \includegraphics[ width=60mm,scale=0.5]{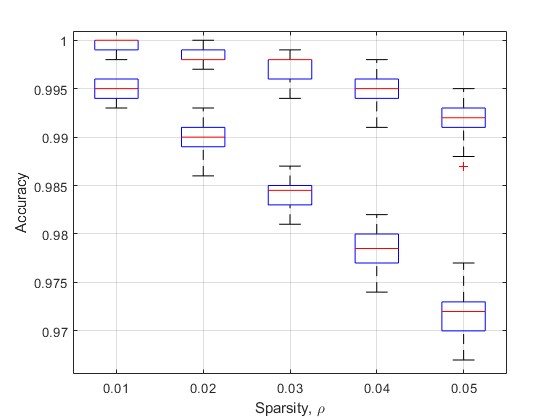}\includegraphics[ width=60mm,scale=0.5]{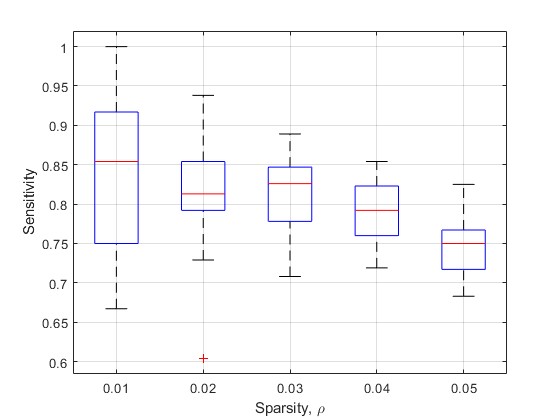}
    \caption{Left: Plot of $C_0$ (lower) and $C_{\pm 1}$ (upper) accuracy measures against signal sparsity $\rho$. Right: Plot of Sensitivity against $\rho$. Note that specificity (True Negative Rate = TN/N) is $\geq 0.999$ for $\rho \in [0.01,0.05]$ so is not plotted. The state space is $\mathcal{S} \in \{20,\ldots,30\}$. Simulations are run 30 times and the red lines display the median value, the boxes cover the interquartile range, the whiskers show the extreme values and outliers are shown with red '+' symbols. The signal sparsity/prevalence range (x-axis) is chosen to represent typical infectious diseases values. The measurement ratio $\alpha$ is constant at $M/N=800/2400$ and no measurement noise is added ($\eta=0$, see discussion in Sec.~\ref{sec:accuracy_noise}). These plots correspond to the experiment  described in Sec.~\ref{sec:accuracy}.}
\label{fig:accuracy_vs_sparsity}
\end{figure}

We carry out numerical experiments with a fixed measurement ratio $\alpha=M/N=800/2400 = 0.33$ and vary the signal sparsity in the range $\rho=[0.01, 0.05]$ to represent typical infectious disease values. 30 simulations are run for each $\rho$ value with $C_0$ and $C_{\pm 1}$ accuracy, true/false positive/negative measures recorded. The high accuracy values shown in Fig.~\ref{fig:accuracy_vs_sparsity} mask an underlying issue with the false negative rates hence our focus on sensitivity. 

\subsection{Accuracy versus measurement noise}
\label{sec:accuracy_noise}
\begin{figure}
    \centering    \includegraphics[ width=80mm,scale=0.5]{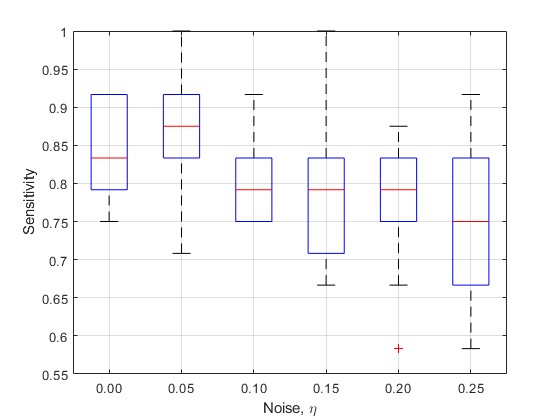}
    \caption{\cut{Plot of }Sensitivity against noise level $\eta$ defined in \eqref{eq:noise_case_2}. This corresponds to experiment  described in Sec.~\ref{sec:accuracy_noise}. Parameters $\rho=0.01, \alpha = M/N = 800/2400$ are fixed. Simulations are run 30 times. The necessity of more iterations for the noisy signals is well known \cite{mezard2008group} so it was necessary to relax the convergence threshold to $10^{-3}$. }
\label{fig:sens_vs_noise}
\end{figure}

In previous sections, we have accounted for the discretization noise arising from the PCR doubling mechanism. Physical noise sources, also inherent in recording $t^y$ values, include fluorescence not being distinguished from background levels e.g. light leaks into the sample well 
and imperfect doubling during the amplification phase e.g. $(1+q)^t$ where $q \in (0,1)$ rather than $2^t$~\cite{ghosh2021compressed}. These tend to lower and raise the measured $C_t$ value from its true reading respectively. In the absence of accurate statistics, we naively assume under- and over- $C_t$ estimation is equiprobable leading to a noisy measurement $\Tilde{t}^y$:
\begin{equation}
\label{eq:noise_case_2}
    p(\Tilde{t}^y) = (1-\eta) ~\delta(\Tilde{t}^y-t^y) + \half \eta  ~\delta\left[(\Tilde{t}^y-t^y)-1\right] + \half \eta~  \delta\left[(\Tilde{t}^y-t^y) + 1\right]
\end{equation}
where $\eta \ll 1$. Setting $\eta=0$ represents no physical measurement noise (which was the case for our previous experiments). Equation \eqref{eq:noise_case_2} is used to add noise to our synthetically generated signals. After convergence of our message passing equations, the inferred $C_t$ values are converted into binary variables where negative corresponds to $C_t = U$ and positive otherwise. 
The resulting sensitivity values, plotted in Fig.~\ref{fig:sens_vs_noise}, assess how robust the inference is to increasing $\eta$. The presence of overlapping tests i.e. each patient participating in multiple tests ($L=KN/M>1$) was considered in~\cite{sakata2020bayesian} and was found to overcome typical noise levels. 

In addition to device-specific measurement noise, modelled by \eqref{eq:noise_case_2}, laboratory sample handling errors can occur \cite{courtney2021using}. These include cross-contamination of samples and errors pipetting samples into physical groups/pools. This source of error is not accounted for in our model. 

\section{Discussion}
\label{sec:discussion}
Pooled testing was originally designed for the efficient diagnosis of infectious diseases~\cite{dorfman1943detection}. There are numerous explanations for the lack of widespread adoption in public health including perceived complexity, requirement for sophisticated laboratory management or conflicting economic incentives. By combining pooled testing efficiencies with accurate estimation of a realistic range of viral loads, the methods presented here may lead to wider adoption into applications such as food contaminant testing and community screening for infectious diseases. 

It is known that the viral load present in infectious disease testing samples can vary by many orders of magnitude  \cite{pan2020viral}. This concept has typically not been taken into account in theoretical investigations of pooled testing. We estimate the viral load by mapping the inference task to a message passing problem where the factor nodes incorporate the specific PCR mixing protocol. Numerical experiments explicitly highlight the source of the error originating from mixing patient samples and we have dealt with this through a modified noise function $\phi(\bt, \ty)$. This filter, used in Sec.~\ref{sec:step_function}, approximates the distributions of random variables $X \equiv \frac{1}{K} \sum\limits_{k=1}^K 2^{-t_k}$ and $Y \equiv 2^{-t^y}$ as step functions with the overlap providing weights in the message passing equations. For completeness, we note that for applications where the viral load actually has a narrow range of values, the well-developed theory of compressed sensing \cite{krzakala2012probabilistic}\cite{krzakala2012statistical} can be applied to infer signal values, accurately and efficiently, using either random or structured pooling matrices. 

The main focus of future work is to improve the scalability of algorithm. While real-valued signals can be approximated via the first two moments of the message distributions, the approximation for our integer-valued state space problem is not clear. The resulting summation in the factor-to-variable messages~\eqref{eq:factor_to_variable_PCR} does not scale well with $|\mathcal{S}|$ (increasing $C_t$ range) or $K$ (number of patients in each pool). Our choice of noise function mitigates this effect by materially reducing the number of summands in the factor-to-variable messages. An alternative approach is to translate the integer $C_t$ values to broad classes of infection status e.g. \cite{ben2022recovery}. Further code efficiencies or parallelization may also help. 

Other avenues of future work include a wet laboratory experiment to test our modelling assumptions. The methods described in this paper can be applied to other amplification protocols such as Loop-mediated isothermal amplification (LAMP). 
\color{black}

\section{Acknowledgments}
This work was funded by the Engineering and Physical Sciences Research Council through grant EP/W015412/1. The authors would like Professor Andrew Beggs for an earlier collaboration on group testing for the SARS-CoV-2 diagnosis.

\section*{References}
\bibliographystyle{unsrt}
\bibliography{bib_BVS}

\end{document}